\documentclass[%
reprint,
nofootinbib,
amsmath,amssymb,
aps,
]{revtex4-2}

\usepackage{graphicx} 
\usepackage{dcolumn} 
\usepackage{bm} 
\usepackage{hyperref} 
\usepackage{color}
\usepackage{soul}
\usepackage[utf8]{inputenc}
\usepackage{enumitem}
\usepackage{dsfont}
\usepackage[dvipsnames]{xcolor}
\usepackage{physics}
\usepackage{makecell}
\usepackage{upgreek}
\usepackage{lipsum, babel}
\usepackage{indentfirst}
\usepackage{fdsymbol}
\usepackage{ulem} 
\graphicspath{ {./} }

\begin{document}

\title{Large-amplitude diamond optomechanics by traversing a nonlinear attractor}

\author{Peyman Parsa, Waleed El-Sayed, Parisa Behjat, Shabir Barzanjeh, and Paul E.\ Barclay}
\email[Paul~E.\ Barclay: ]{Corresponding author pbarclay@ucalgary.ca}
\affiliation{
Institute for Quantum Science and Technology, University of Calgary, Calgary, Alberta T2N 1N4, Canada}

\date{\today}

\begin{abstract}
Nonlinear dynamics clamp the amplitude of mechanical resonators driven into self-oscillation by optomechanical backaction. Here we overcome the conventional limits of self-oscillation amplitude by navigating the nonlinear dynamical landscape of a diamond optomechanical cavity supporting coherent optomechanics at room temperature. By exploiting the bistable phase space of the system, we increase the oscillation amplitude by nearly an order of magnitude.  This enhancement arises from deterministic access to a high-energy state in the system's nonlinear attractor, and is accompanied by the generation of an optical frequency comb produced by cascaded phonon scattering that underlies the nonlinear dynamics. Our results establish nonlinear attractor engineering as a route to large-amplitude coherent phonon generation and provide a platform for optomechanical frequency combs, spin–mechanical interfaces in diamond, and precision sensing in ambient conditions.
\end{abstract}

\maketitle

\textit{Introduction}
The radiation pressure force exerted by light on mechanical resonators allows their motion to be sensitively controlled, and underlies technologies ranging from nanoscale sensing \cite{Krause2012, PhysRevX.4.021052} to gravitational wave detection \cite{Abbott2016, Page2021} and quantum information processing \cite{barzanjeh2022optomechanics}. The ability of Mechanical resonators ability to interact with light and many other physical systems,  makes optomechanics promising for interfacing quantum technologies possessing no intrinsic optical coupling such as superconducting circuits \cite{ref:Regal2011} and spin qubits \cite{Peng2025}. Diamond optomechanical devices excel in this regard \cite{Shandilya:22}, thanks to diamond's ability to host spin qubits that can be coupled to mechanical resonators, enabling spin-mechanical \cite{ref:arcizet2011snv, ref:macquarrie2013msc, ref:ovartchaiyapong2014dsc, ref:teissier2014scn, ref:lee2017topical} and spin-optomechanical \cite{Shandilya:22, ref:delord2020spin} interfaces. Diamond nanomechanical devices can also store and manipulate \cite{lake2021processing} optical signals and to modify the properties of spin qubits \cite{ref:kuruma2025controlling, ref:joe2025observation}.

The dynamics of cavity optomechanical systems are governed by Hamiltonian $H_\text{int} = -\hbar g_0a^\dagger a \left(b + b^\dagger\right)$, where $g_0$ is the coupling rate between a single optical cavity photon and a single mechanical resonator phonon, whose annihilation (bosonic) operators are $a$ and $b$, respectively. Although this interaction leads to nonlinear Heisenberg equations of motion, when linearized it can explain many optomechanical phenomena, such as resonator damping and anti-damping from optomechanical backaction in systems whose resonator frequency, $\omega_{\text{m}}$, is higher than the optical cavity's energy decay rate, $\kappa$ \cite{PhysRevLett.103.103601}. However, linear treatments fail when mechanical oscillations are large \cite{Navarro-Urrios2017}, as in backaction driven self-oscillations required by applications such as optomechanical spin control \cite{Shandilya2021}, bistable memories \cite{ref:bagheri2011dmn}, frequency comb generation \cite{PhysRevLett.127.134301}, and optomechanical sensors \cite{PhysRevApplied.14.024079, PhysRevApplied.16.044007}.  

Of particular importance for these applications are limits imposed by nonlinear dynamics on the achievable self-oscillation amplitude. These limits arise from amplitude-dependent dynamical backaction, which reshapes the mechanical gain profile and creates multiple stable limit cycles in the system phase space. Poot et al.\ \cite{ref:poot2012bls} analyzed this nonlinear regime and showed that the attainable oscillation amplitude depends on the ratio $\omega_\mathrm{m}/\kappa$, which controls the strength of cavity dynamical backaction. In sideband unresolved systems ($\omega_\text{m}/\kappa \ll 1$), successive Stokes scattering processes, referred to here as cascaded phonon scattering, have been used to enhance self-oscillation amplitude \cite{ref:bagheri2011dmn}, while in sideband resolved systems ($\omega_\text{m}/\kappa \gg 1$) optomechanical bistability has been harnessed to switch between stable self-oscillating states   \cite{PhysRevLett.115.233601}.
Here we combine these concepts to enhance self-oscillations of a optomechanically driven diamond resonator. We access a large amplitude state of its bistable dynamics where cascaded phonon scattering is maximized, and use heterodyne spectroscopy to directly measure frequency comb signatures of this scattering process, demonstrating its importance despite the device's nearly sideband resolved nature ($\omega_\text{m} \sim \kappa$). Finally, by measuring the system's nonlinear attractor we show that it can function as a slow force-sensor with 40 dB amplification. This behavior is observed in ambient conditions, a critical step forward from studies with silicon devices requiring cooling to cryogenic temperature \cite{ref:safavinaeini2012oqm}, thanks to the diamond system's high optomechanical cooperativity $C_\text{om} = 4N g_0^2/\kappa\Gamma_\text{m}$, arising from its large $g_0$, low optical ($\kappa$) and mechanical ($\Gamma_\text{m}$) dissipation, and ability to support large intracavity photon number $N$ without inducing nonlinear absorption. These advances enable optomechanical control of diamond spin qubits and may allow access to the coherent regime through a predicted twentyfold increase in resonator–spin coupling relative to previous demonstrations \cite{Shandilya2021}.

\begin{figure}
    \centering
    \includegraphics[width = \linewidth]{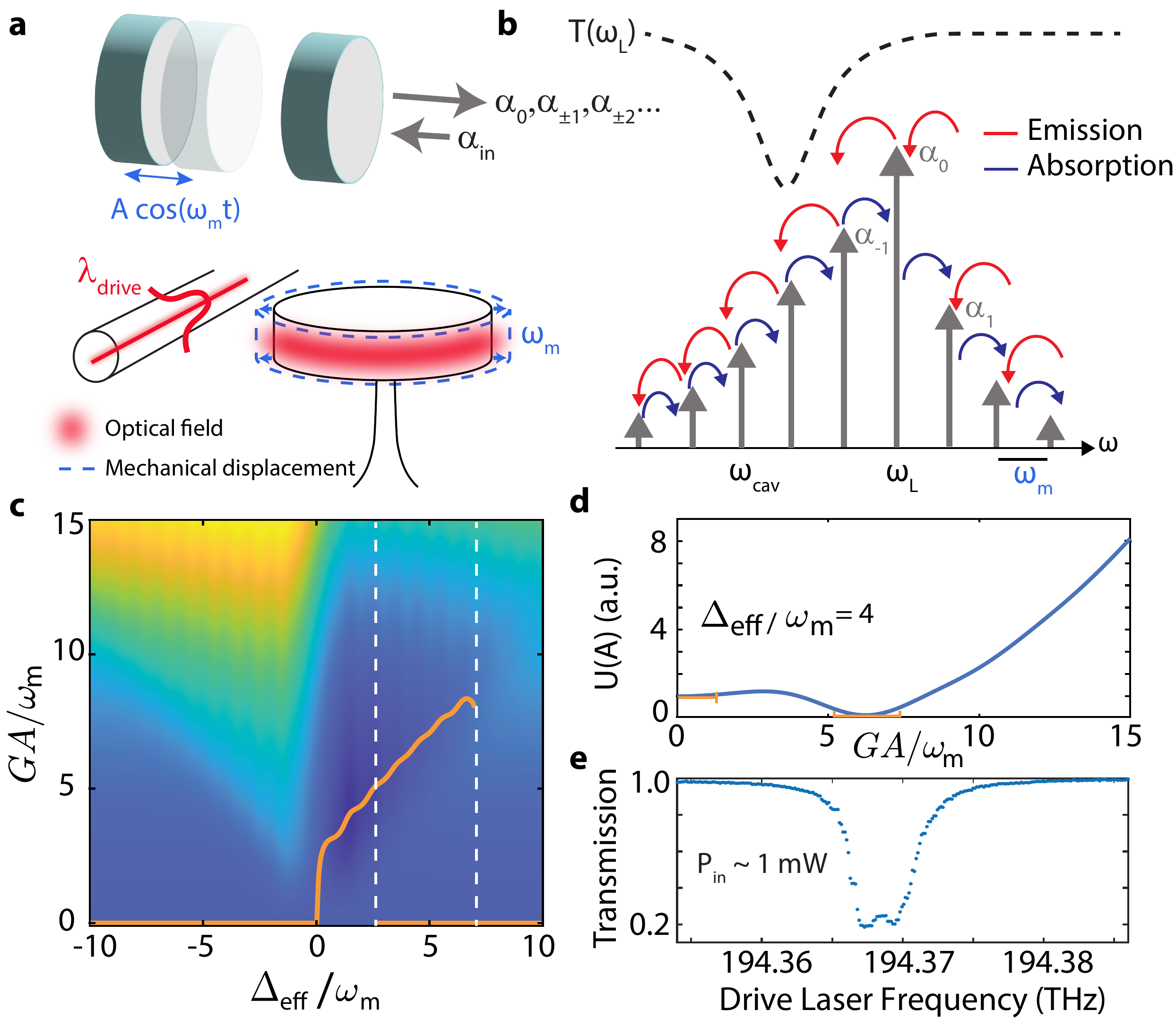}
    \caption{(a) Cavity optomechanical systems. Top: Fabry-P\'{e}rot cavity whose mirror oscillations create optical sidebands with amplitude $\alpha_\text{n}$. Bottom: the fiber taper coupled microdisk system studied here. (b) Scattering of the cavity field into sidebands is enhanced when they are on-resonance with the cavity mode. (c) The diamond microdisk's effective potential $U(A)$ as a function of normalized mechanical amplitude and drive laser detuning from resonance. The orange lines trace local minima of $U$. Mechanical bistability is possible in the region between the dotted lines. (d) Effective potential at a fixed detuning, with local minima highlighted. (e) Drive laser transmission for low input power.}
    \label{fig:fig1}
\end{figure}

\textit{Background} \label{phononLasing}
Optomechanical damping and anti-damping is central to demonstrations of optomechanical cooling  \cite{ref:chan2011lcn, ref:teufel2011scm} and phonon lasing \cite{PhysRevLett.94.223902, Rokhsari:05, PhysRevLett.95.033901, PhysRevLett.115.233601}, respectively. This latter self-oscillation phenomena corresponds to excitation of a stable mechanical limit cycle \cite{PhysRevLett.115.233601}, and is normally achieved when the input laser is blue-detuned from the cavity resonance by the mechanical resonator frequency, $\Delta = \omega_\text{m}$, and  $C_\text{om} > 1$. In a generic cavity optomechanical system, illustrated in Fig.\ \ref{fig:fig1}(a), these dynamics are captured by equations of motion \cite{aspelmeyer2014cavity}
\begin{align}
    &\dot{\alpha} = -\frac{\kappa}{2} \alpha + i(\Delta_\text{eff} + Gx)\alpha +\sqrt{\kappa_\text{ex}}\alpha_\text{in},\\
    &\ddot{x} + \Gamma_\text{m}\dot{x} + \omega_\text{m}^2 x = \frac{\hbar G}{m_\text{eff}}|\alpha|^2,
    \label{eq:eq854}
\end{align}
where $x$ is the mechanical resonance displacement, $m_\text{eff}$ is its effective mass, $\alpha = \langle a \rangle = \sqrt{N}$ is the amplitude of the optical cavity mode, and $G$ is the optomechanical coupling coefficient related to $g_0= G x_\text{zpf}$ by the resonator's zero point fluctuation amplitude $x_\text{zpf} = \sqrt{\hbar/2\omega_\text{m}m_\text{eff}}$. For a given field $\alpha_\text{in}$ (photon flux $|\alpha_\text{in}|^2$) input to the cavity mode, the output field is $\alpha_\text{out} = \alpha_\text{in} - \sqrt{\kappa_\text{ex}}\alpha$. Here $\kappa_\text{ex}$ is the coupling rate between the cavity and the input-output channel, which for integrated photonic cavities is typically a waveguide. Note that $\Delta_\text{eff} = \Delta + \delta\Delta_\text{abs}$ includes any quasi-static shifts $\delta\Delta_\text{abs}$ to the cavity resonance frequency due to photothermal effects.

The motion of a self-oscillating resonator can be described by $x(t) = x_0 + A \cos(\Omega_\text{m} t)$, where $A$ is oscillation ampltiude, $x_0$ is the average displacement and $\Omega_\text{m}$ is the frequency of oscillations and differs slightly from $\omega_\text{m}$ due to the optical spring effect ($\Omega_\text{m} - \omega_\text{m} = \delta \omega_\text{m}$). 
%
This satisfies the equations of motion when $\alpha(t) = e^{i\beta \sin(\Omega_\text{m} t)} \sum_{n=-\infty}^{\infty} \alpha_\text{n} e^{-in\Omega_\text{m} t}$, where $\beta = GA/\omega_\text{m}$ is the optomechanical modulation index that together with $\kappa$ determines the sideband amplitudes
\begin{equation}
    \alpha_\text{n} \approx \frac{\sqrt{\kappa_\text{ex}}\alpha_\text{in}J_n(\beta)}{\frac{\kappa}{2} - i(\Delta_\text{eff}+n\Omega_\text{m})}.
    \label{eq:eq6234}
\end{equation}
For sufficiently large modulation index $\beta$, higher-order sidebands contribute significantly to the intracavity field, enabling cascaded phonon-scattering processes that redistribute optical power across the comb spectrum. As illustrated in Fig.\ \ref{fig:fig1}(b), these sidebands are enhanced when they are resonant with the cavity mode. In this regime the mechanical motion phase-modulates the intracavity field and generates higher-order Stokes and anti-Stokes sidebands spaced by integer multiples of the mechanical frequency $\omega_m$.
The corresponding mechanical resonator amplitude $A$ satisfies 
\begin{equation}
    \begin{gathered}
        \frac{dA}{dt} = -\frac{\Gamma_\text{m}}{2m_\text{eff}\omega_\text{m}^2} \frac{\partial U(A)}{\partial A},
    \end{gathered}
    \label{eq:eq421}
\end{equation}
where $U(A)$ is the effective potential defined by
\begin{equation}
    U(A) = \frac{1}{2}m_\text{eff}\omega_\text{m}^2 A^2 - \frac{2}{\Gamma_\text{m}} \int_{0}^{A} \text{P}_\text{rad}(A') \frac{dA'}{A'}.
    \label{eq:eq871235}
\end{equation}
in which the allowed values of $A$
occur at local minima of $U(A)$, each representing a stable limit cycle of the mechanical oscillator, i.e., a dynamical attractor, while the barriers separating the minima set the thresholds for transitions between attractors \cite{PhysRevLett.96.103901}.  For large $A$, many sidebands can contribute to the optical power transferred to the resonator $\text{P}_\text{rad} (A) = \hbar G \left\langle |\alpha|^2\dot{x} \right\rangle = \hbar G \omega_\text{m} A  \sum_{n = -\infty}^{n = \infty} \text{Im}{\{ \alpha_\text{n}\alpha_\text{n+1}^*\}}$, and as illustrated in Fig.\ \ref{fig:fig1}(b) and shown below, $A$ can be enhanced through cascaded phonon scattering that generates multiple phonons from a single input photon.

\textit{Experiment}
Our nonlinear cavity optomechanics platform is a diamond microdisk that allows coherent ($C_\text{om}>1$) optomechanics to be accessed at room temperature \cite{lake2020two, lake2021processing}. The device studied here was characterized in  Ref.\ \cite{ref:behjat2023kos} and supports a mechanical radial breathing mode with frequency $\omega_\text{m}/2\pi \approx 2.31$ GHz and quality factor $Q_\text{m} \approx 9,000$ that is coupled to an optical whispering gallery mode with frequency $\omega_\text{o} \approx 2\pi \times 194~\text{THz}$ (wavelength $\lambda_\text{o} \approx 1543$ nm) and quality factor $Q_\text{o} \approx 70,000$.  
In Fig.\ \ref{fig:fig1}(c) we plot the effective potential $U(A)$ of this device as a function of input laser detuning.
Stable mechanical self-oscillation is predicted at points traced by the contours in Fig.\ \ref{fig:fig1}(c), and is predicted over a wide range of blue detuning. For some of this range, the mechanical resonator exhibits bistability ($A = 0$ or finite amplitude) as highlighted by the plot of $U(A)$ for fixed $\Delta_\text{eff} = 4\ \omega_\text{m}$ in Fig.\ \ref{fig:fig1}(d). 

We optically excite and measure self-oscillations by evanescently coupling light into and out of the microdisk with a fiber taper waveguide. 
The drive laser couples to the mode at $\omega_\text{o}$ and excites the resonator through optomechanical anti-damping. To independently monitor resonator motion, a probe laser is coupled to a lower $Q_\text{o}$ mode (wavelength 1507 nm) whose linewidth is much larger than $\omega_\text{m}$. The probe and drive fields are separated using an optical filter at the waveguide output before being detected. 
The lineshape of the time averaged drive mode transmission $T = \overline{|\alpha_\text{out}/\alpha_\text{in}|}^2$ is shown in Fig.\ \ref{fig:fig1}(e) for low input power $P_\text{in} < 1\,\text{mW}$. Note that this mode is a doublet due to backscattering within the microdisk \cite{ref:kippenberg2002mct}; this is taken into account in all of our calculations.

\begin{figure}
    \centering
    \includegraphics[width = \linewidth]{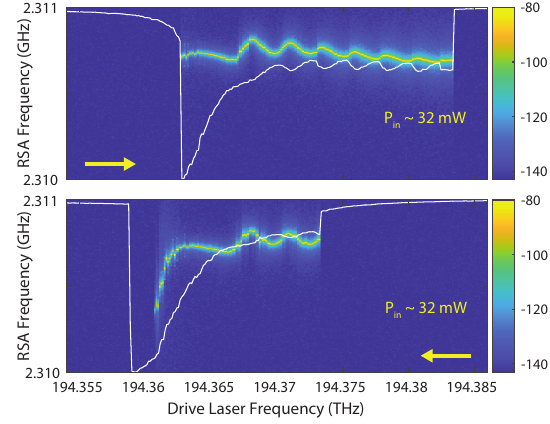}
    \caption{Drive field transmission lineshape for high input power (white line) and corresponding spectrograph of the transmitted probe intensity's power spectral density over the frequency range of the mechanical radial breathing mode, when the drive laser detuning is swept from red to blue (top) and blue to red (bottom). Color scale has units $\text{dB}/\sqrt{\text{Hz}}$.} 
    \label{fig:fig2}
\end{figure}

To induce mechanical self-oscillations, we increase $P_\text{in}$ to $32\,\text{mW}$. The drive mode lineshape at this high power, shown in Fig.\ \ref{fig:fig2}, exhibits two distinct features in the blue detuned region: wave-like oscillations, and hysteresis. Both of these effects are related to self-oscillation of the resonator, as revealed by the spectrographs in Fig.\ \ref{fig:fig2} that show, for varying $\Delta$, the power spectral density near $\omega_\text{m}$ of the photodetected probe field. A sharp peak generated by self-oscillations is visible in the spectrographs, and its amplitude and frequency exhibit similar hysteretic behaviour as the drive laser transmission. The drive mode also exhibits hysteresis near the resonance from commonly observed photothermal effects \cite{ref:carmon2004dtb} unrelated to optomechanical dynamics.

\begin{figure*}
  \centering
\includegraphics[width = \linewidth]{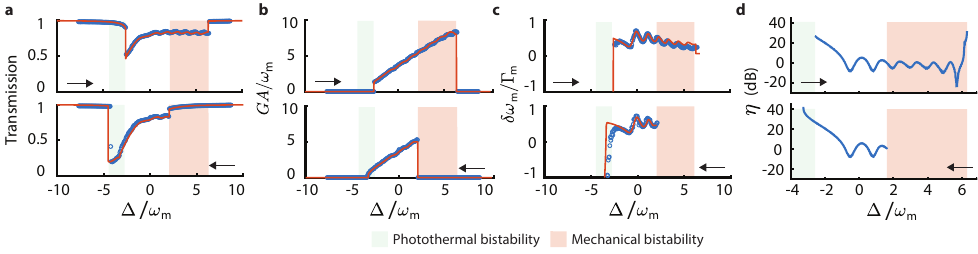}
    \caption{Measured (blue) and predicted (red) effect of drive laser detuning on (a) drive laser transmission, (b) self-oscillation amplitude $A$ transduced by the probe laser, (c) mechanical self-oscillation frequency, and (d) signal gain $\eta$ derived from $GA(\Delta)$.}
\label{fig:fig2b}
\end{figure*}

\textit{Optomechanical Bistability}
The observed hysteresis in $A$ and $\omega_\text{m}$ are plotted as a function of $\Delta$ in Fig.\ \ref{fig:fig2b}. Also shown are corresponding theoretical predictions, obtained by fitting the drive laser transmission in Fig.\ \ref{fig:fig2} with  $T(\Delta)$ from Eq.\ \eqref{eq:eq854}. 
To reproduce the experimental data, a static photothermal shift was included in the effective detuning, $\Delta_\text{eff} = \Delta + \zeta_\text{TO} N$, where $\zeta_\text{TO}$ is the frequency shift per photon due to the thermo-optic effect.  
The resulting plots show good agreement between measurement and theory, and highlight two regions of bistability. Purely mechanical bistability, which is independent of photothermal effects, is present in the blue detuned region. Photothermal bistability is present near resonance; this affects the mechanical resonator indirectly through its influence on $N$, but as mentioned above, is not optomechanical in nature. 
In the mechanical bistability region, the device jumps between $A = 0$ and finite $A$. This switching reflects the coexistence of two stable attractor states in the nonlinear optomechanical phase space. Notably, when scanning the drive laser from low to high frequency in this region, $A$ grows monotonically to $A \sim 9\ \omega_{\text{m}}/G$ for $\Delta \sim 6\,\omega_\text{m}$. Conversely, when scanning from high to low frequency, self-oscillations begin closer to resonance ($\Delta \sim 3\omega_{\text{m}}$) and have a lower maximum $A$. 

The above measurements show that by navigating the device's nonlinear dynamics, a 6 times enhancement to $A$ can be achieved compared to operating at the conventional self-oscillation detuning ($A \sim 1.5\ \omega_{\text{m}}/G$ at $\Delta = \omega_\text{m})$. This enhancement is of significant interest for implementing spin-mechanical control using large dynamical stress fields within the cavity since the strain generated by the mechanical motion scales linearly with the oscillation amplitude \cite{ref:macquarrie2013msc, Shandilya2021}. Assuming $G/2\pi = 78\,\text{GHz/nm}$, as characterized in previous studies \cite{ref:behjat2023kos}, we predict a maximum amplitude of $A|_\text{max} \sim 177 \,\text{pm}$, which is nearly 20 times larger than in previous work \cite{Shandilya2021} and is sufficient to enable coherent optomechanical spin manipulation \cite{ref:macquarrie2013msc}.

Navigating the resonator's attractor also allows its potential as a force sensor to be enhanced. Self-oscillating resonators can sense forces via tracking resonator frequency \cite{PhysRevApplied.14.024079}, sideband strength \cite{Javid2021}, and oscillation amplitude. 
In the latter approach, sensitivity to an external force $F_\text{ex}(t)$ is governed by its effect on detuning $\delta\Delta(t) \approx \frac{G}{m\omega_\text{m}^2} F_\text{ex} (t)$ and change in $A$,
\begin{equation}\label{eq:ampl}
    \delta A (t) = \frac{1}{G}\left(\frac{\partial (GA)}{\partial \Delta_\text{eff}}\right) \delta \Delta(t),
\end{equation}
where we have assumed that $|\delta\Delta| \ll \omega_\text{m}$ and that changes are slow compared to $\omega_\text{m}$ and $\Gamma_\text{m}$.
If the force is modulated with frequency $\Omega$, $\delta(t)$ will be optomechanically transduced into sidebands spaced by $\Omega$. 
Equation \eqref{eq:ampl} shows that the signal gain at a given operating condition, $\eta = \left(\frac{\partial (GA)}{\partial\Delta_\text{eff}}\right)^2$, is largest for points on the attractor diagram with the highest slope \cite{PhysRevLett.96.103901}. This is reflected by Fig.\ \ref{fig:fig2b}(d), which plots $\eta(\Delta)$ derived from the measured $GA(\Delta)$ in Fig.\ \ref{fig:fig2b}(b), and shows that $\eta$ increases significantly near the transition points at the edges of the attractor, exhibiting over 40 dB of enhancement. This amplification arises from the steep slope of the nonlinear attractor landscape, which converts small external perturbations into large variations in oscillation amplitude. Note that near the low (high) frequency boundary, $\eta$ is larger when scanning from high to low (low to high) than vice versa due to the system's phototermal (mechanical) bistability. 
Finally, it is critical to emphasize that while $\eta$ amplifies thermal and other fundamental noise sources affecting resonator motion, it reduces the relative strength of technical noise, thus providing an overall boost in signal to noise for systems limited by detector noise.

\begin{figure}
    \centering
    \includegraphics[width = \linewidth]{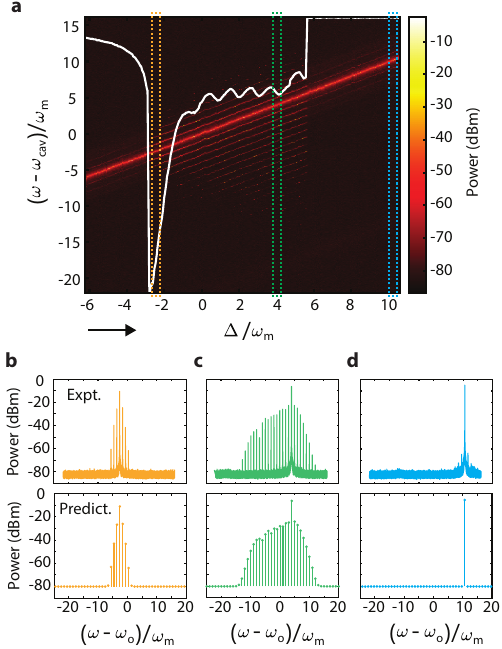}
    \caption{(a) Optical spectrum of the transmitted drive field for varying detuning. The corresponding average drive field transmission is shown in white. (b) Measured (top) and predicted (bottom) spectra at fixed detunings indicated by dashed lines in (a). }
    \label{fig:fig3}
\end{figure}

\textit{Cascaded Brillouin Scattering}
For both drive laser sweep directions we observe oscillations in $T(\Delta)$ spaced by $\Omega_\text{m}$.  These features arise from high frequency modulation of the output field by self-oscillations, resulting in a change in the time averaged transmission. They indicate that additional scattering processes driving self-oscillations are activated at these detunings, and are reproduced by theoretical plots of $T(\Delta)$ in Fig.\ \ref{fig:fig2b}(a). 
The relationship between $\Delta$ and scattered sidebands can be directly verified by measuring the optical spectrum of the transmitted drive field, which we obtained using a heterodyne spectrometer (APEX)  and plot in Fig.\ \ref{fig:fig3}(a). The dominant peak in each spectrum is at the drive laser frequency. Over the range of $\Delta$ where the resonator self-oscillates, sidebands at multiples of $\Omega_\text{m}$ emerge with height  $\propto |\alpha_{\text{n}}(\Delta)|^2$.  In Figs.\ \ref{fig:fig3}(b - c), we compare theoretical predictions of the sideband intensities with measured spectra for fixed values of $\Delta$ near the edges of the self-oscillation region, and observe excellent agreement. From the measured spectra we see that the number of sidebands increases as $\Delta$ becomes more blue detuned; this is consistent with the data in Fig.\ \ref{fig:fig2b}(b) that shows that $A$, and as a result $|\alpha_\text{n}|^2 \propto |J_\text{n}(\beta)|^2$, is maximum near the blue detuned edge of the self-oscillation region. While this is consistent with predictions in Fig.\ \ref{fig:fig2b}, it is counterintuitive given that the drive laser is far from cavity resonance at this large detuning.  It is also in contrast to cascaded phonon generation in sideband unresolved systems \cite{ref:bagheri2011dmn}, where scattering between sidebands falls within the cavity's optical linewidth.  We highlight that the large amplitude state is only accessible by traversing the bistable nonlinear attractor from red to blue detuning. 
We also emphasize that the amplitude enhancement is optomechanical in nature, and is not driven by photothermal effects. Finally, we note that closely spaced peaks visible in our data are related to laser phase noise sidebands previously discussed in Ref.\ \cite{Kippenberg_2013}, which are shown by the laser's optical spectrum in Fig.\ \ref{fig:fig3}(d) to occur at $\approx$ 2.3 GHz, almost coinciding with $\omega_\text{m}$. 

The cascaded phonon generation process can be framed as arising from a Brillouin scattering interaction described by the Hamiltonian $H = \hbar \omega_\text{cav}\sum_{n = -\infty}^{\infty}a_\text{n}^\dagger a_\text{n} + \hbar \omega_\text{m}b^\dagger b - \hbar g_0 \left[\sum_{n = -\infty}^{\infty} a_\text{n+1}^\dagger a_\text{n}b + H.C.  \right]$, with $a_\text{n}$ being the annihilation operator of each sideband. 
In a self-oscillating resonator with mean phonon number $n_\text{SO}$, the dominant interaction is given by the Hamiltonian $H_\text{interaction} = -\hbar g_0 \sqrt{n_\text{SO}}e^{-i\omega_\text{m}t}\sum_{n = -\infty}^{\infty} a_\text{n+1}^\dagger a_\text{n} + H.C.$ This gives the unitary evolution operator 
\begin{equation}
S = \exp \left[i\frac{g_0\sqrt{n_\text{SO}}}{\omega_\text{m}}\left(\sum_{n = -\infty}^{\infty} a_\text{n+1}^\dagger a_\text{n} + H.C. \right)\right],
\end{equation}
which is the scattering operator of a phase modulator \cite{Capmany:10}.
Although this analysis neglects dissipation, it highlights that interactions between sidebands are limited to nearest neighbours. This is due to 2-phonon transitions rates scaling with the second power of $g_0/\omega_\text{m}$, which is small in most cavity optomechanical systems. However, in the absence of direct multi-phonon processes, each cavity photon can emit (absorb) a phonon and become a new photon with a frequency decreased (increased)  by $\Omega_\text{m}$. Through this mechanism, cascaded interactions between sidebands and an enhancement to resonator amplitude are realized. The resulting optical spectrum forms a frequency comb whose spacing is fixed by the mechanical oscillation frequency.

\textit{Conclusion}
Our measurements have demonstrated the potential for exploiting the nonlinear dynamics of a bistable cavity optomechanical system to to achieve large oscillation amplitude. 
The cascaded phonon scattering process underlying these dynamics was studied using heterodyne measurements for the first time, which confirmed that a 6 times enhancement in self-oscillation amplitude can be achieved when operating in the far blue detuned regime of the bistable device's nonlinear dynamical landscape. The resulting enhanced resonator amplitude  will play an important role in frequency comb generation \cite{PhysRevLett.132.163603, PhysRevLett.114.013601} and optomechanical control of spin qubits \cite{Shandilya2021} and enables exploration of nonlinear optomechanical regimes beyond the conventional self-oscillation limit \cite{Brawley2016, PhysRevA.93.033846, Navarro-Urrios2017, PhysRevLett.132.053601}. In addition, when operated near the zero-detuned boundary of self-oscillation, this device can provide 40 dB of amplification of signals generated by slow external forces. In future, this system can be used to extend studies of the auto-correlation of a phonon laser in the linear regime \cite{Cohen2015} to include cross-correlation of higher order sidebands. 

\bibliography{nano_bib}

\end{document}